\pdfoutput=1
\documentclass{JINST}

\title{The UA9 experimental layout}

\author{W. Scandale$^{1,2,6}$, G. Arduini$^1$, R. Assmann$^1$, C. Bracco$^1$, F. Cerutti$^1$, J. Christiansen$^1$,
S. Gilardoni$^1$, E. Laface$^1$,  R.Losito$^1$,  A. Masi$^1$,  E. Metral$^1$,  D. Mirarchi$^1$, S. Montesano$^1$, 
V. Previtali$^1$,  S. Redaelli$^1$, G. Valentino$^1$, P. Schoofs$^1$, G. Smirnov$^1$,  L. Tlustos$^1$,
E. Bagli$^3$, S. Baricordi$^3$, P. Dalpiaz$^3$, V. Guidi$^3$, A. Mazzolari$^3$,  D. Vincenzi$^3$, 
S. Dabagov$^4$, F. Murtas$^4$,
A. Carnera$^5$, G. Della Mea$^5$,  D. De Salvador$^5$,  A. Lombardi$^5$,  O. Lytovchenko$^5$,  M. Tonezzer$^5$, 
G. Cavoto $^{6}$\thanks{Corresponding author}, L. Ludovici$^{6}$, R. Santacesaria$^{6}$, P. Valente$^{6}$, F. Galluccio$^{7}$,
A.G. Afonin$^{8}$, M.K. Bulgakov$^{8}$,  Yu.A. Chesnokov$^{8}$, V.A. Maisheev$^{8}$,  I.A. Yazynin$^{8}$, 
A.D. Kovalenko$^{9}$,  A.M. Taratin$^{9}$,  Yu.A. Gavrikov$^{10}$,  Yu.M. Ivanov$^{10}$, 
L.P. Lapina$^{10}$, V.V. Skorobogatov$^{10}$, 
W. Ferguson$^{11}$,  J. Fulcher$^{11}$, G. Hall$^{11}$,  M. Pesaresi$^{11}$, M. Raymond$^{11}$, A. Rose$^{11}$, M. Ryan$^{11}$,  O. Zorba$^{11}$, G. Robert-Demolaize$^{12}$,  T. Markiewicz$^{13}$,  M. Oriunno$^{13}$,  U. Wienands$^{13}$ \\
\llap{$^1$}CERN, European Organization for Nuclear Research, CH-1211 Geneva 23, Switzerland\\
\llap{$^2$}Laboratoire de l'Accelerateur Lineaire (LAL), Universite Paris Sud Orsay, Orsay France\\
\llap{$^3$}INFN Sezione di Ferrara, Dipartimento di Fisica, Universita` di Ferrara, Ferrara, Italy\\
\llap{$^4$}INFN LNF, Via E. Fermi, 40 00044 Frascati (Roma) Italy\\
\llap{$^5$}INFN Laboratori Nazionali di Legnaro, Viale Universita` 2, 35020 Legnaro (PD), Italy\\
\llap{$^6$}INFN Sezione di Roma, Piazzale Aldo Moro 2, 00185 Rome, Italy\\
\llap{$^7$}INFN Sezione di Napoli, Italy\\
\llap{$^8$}Institute of High Energy Physics, Moscow Region, RU-142284 Protvino, Russia\\
\llap{$^9$}Joint Institute for Nuclear Research, Joliot-Curie 6, 141980, Dubna, Moscow Region, Russia\\
\llap{$^{10}$}Petersburg Nuclear Physics Institute, 188300 Gatchina, Leningrad Region, Russia\\
\llap{$^{11}$}Imperial College, London, United Kingdom\\
\llap{$^{12}$}Brookhaven National Laboratories P.O. Box 5000 Upton, NY 11973-5000, USA\\
\llap{$^{13}$}SLAC National Accelerator Laboratory 2575 Sand Hill Road Menlo Park, CA 94025, USA\\
 E-mail: \email{gianluca.cavoto@roma1.infn.it}}

\abstract{
The UA9 experimental equipment  was installed in the CERN-SPS in
March '09 with the aim of investigating crystal assisted
collimation in coasting mode. 
 Its basic layout comprises silicon  bent crystals acting as primary collimators  mounted inside two 
vacuum vessels.  A movable 60 cm long block of tungsten  located 
downstream at about 90 degrees phase advance  intercepts  the deflected beam.
 Scintillators,   Gas  Electron Multiplier chambers    and other 
beam loss monitors measure nuclear loss rates induced by
the interaction of the  beam  halo in the crystal. Roman pots  are installed in the path of the deflected
particles and are 
equipped with a Medipix detector to reconstruct the
transverse distribution of the impinging beam.  Finally UA9 takes
advantage of an LHC-collimator prototype installed close
to the Roman pot to help in setting the beam conditions
and to analyze  the efficiency to  deflect the beam.  This paper describes in details the hardware
installed to  study the crystal collimation  during 2010. }


\begin{document}

\section{\label{sec:intro} Introduction}

  Halo particles in circular accelerator represent a threat for the 
  good performance, stability and protection  of the machine.
   Specific beam  collimation system must be designed and implemented based on passive 
   objects  able to scatter and absorb undesired particles.
   In the last years bent crystals have  been efficiently  used
 to extract beam particles out of an accelerator \cite{Pbextraction,Protvinoext,NA48KS} using 
 the coherent interaction of the charged particles with the crystal ({\it crystal channeling}).
 The crystal extraction can be therefore applied to the main beam but also
 to halo particles\cite{TevatronRHIC}. A crystal assisted collimation system for hadron colliders 
 (as the Large Hadron Collider, LHC) has been  proposed \cite{LHCproposal}.
 
   A classic two-stage collimation system   \cite{opticstwostage}  in accelerators consists of a primary element acting as a small
   scattering target and a secondary element absorbing particles impinging on it. An amorphous primary target
   scatters particles in no preferred direction while a bent crystals traps particles with the coherent scattering on
  aligned atomic planes and kicks them in only one direction. 
   The halo protons  can be redirected  so that they hit the secondary absorber with a large impact parameter and, Êtherefore, can be efficiently removed.

     CERN approved in 2008 the UA9 experiment with the aim of testing
     directly  the crystal assisted  collimation as an alternative for both protons and lead ion beam  collimation in the LHC.
   UA9's final goal is to demonstrate that a crystal based collimation has a higher cleaning efficiency than traditional scheme.
   This paper therefore  focuses on the layout of the
UA9 equipment installed in the SPS ring. Various beam
instrumentation and detectors are  in fact used to measure the beam losses close to the crystal and around the ring with different level 
of accuracy and redundancy. The procedures to detect  and establish the channelling condition  and to  measure the collimation efficiency
are subjects  of other publications.

\section{\label{sec:collregion} The experimental  region}

\begin{figure*}[ht]
\centering
\includegraphics[width=\linewidth]{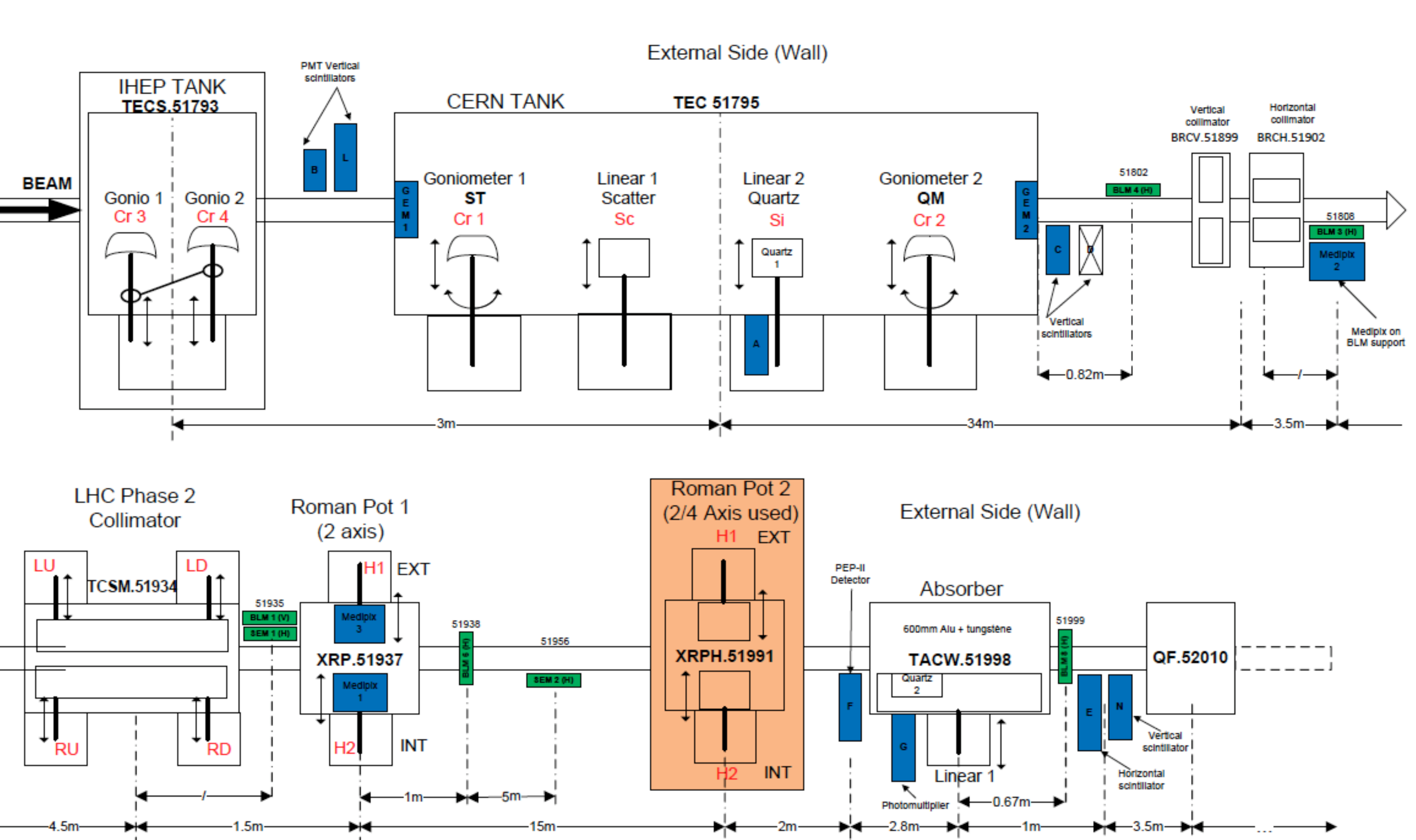}
\caption{UA9 collimation region layout as in 2010 data-taking.}
\label{fig:Layout1}
\end{figure*}

\begin{figure*}[ht]
\centering
\includegraphics[width=\linewidth]{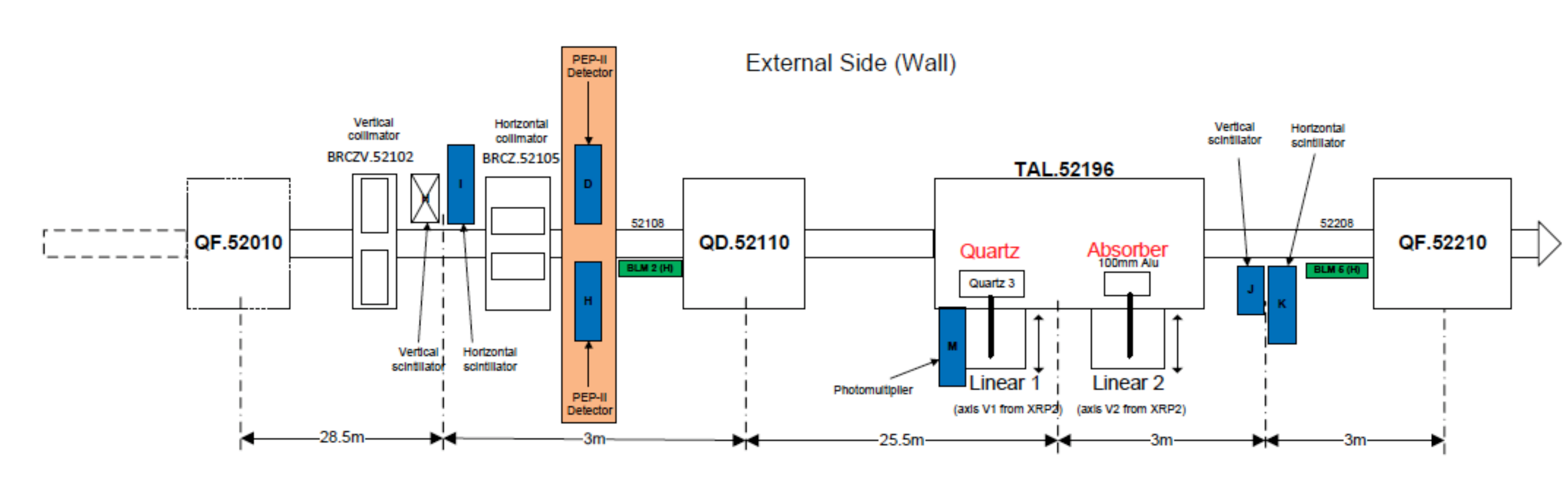}
\caption{External beam dump in UA9 layout.}
\label{fig:Layout2}
\end{figure*}

The UA9 experiment is installed in the Long Straight
Section 5 of the SPS. In this paper we describe the layout that was 
set in place for the 2010 data-taking apart from small  further adjustments of the single components.
 In Fig.\ref{fig:Layout1} the instrumented beam line is shown  where the beam cleaning and the losses should be concentrated (the collimation region). 
 Two vacuum vessels  are equipped with  four silicon crystals mounted on precision
 goniometers. One tank was manufactured at Institute for High Energy Physics (``IHEP tank'') and the   other  belonged to the RD22 experiment and was recently refurbished at CERN (``CERN tank''). Inside the  CERN tank a tungsten scatterer and a Cherenkov detector are mounted to make 
 direct measurement with an amorphous material  and to quantify the number of impinging particles on the crystal 2 (Cr2). 
 Channeled protons are coherently  deflected by the angle  given by the crystal bending angle (around 100 $\mu$rad).  About 60 m downstream  the crystals  (position ``TACW.51998'' )  the  channeled protons are  displaced  by several mm with respect to the primary beam envelope and they are  then collected   on a secondary  tungsten  collimator, instrumented 
 with another Cherenkov detector. 
 In close proximity  and outside of the vacuum vessels some detectors  are installed which are  sensitive to the production of secondary particles
 from inelastic interactions in the obstacles present within the beam pipe. Several scintillators counters equipped with PMTs, Gas Electron Multiplication chambers  (GEM)  
  and LHC-type ionization chamber Beam Loss  Monitor (LHC BLM) are used for this purpose.  
  About 45 m downstream the crystals a Roman Pot device (RP1, at ``XRP.51937'')  is installed.  It comprises two  horizontal axes  (H1 and H2 in Fig.\ref{fig:Layout1}) and each axis supports one  Medipix  \cite{medipixpaper}  pixel detector that can be moved towards the center of the beam to intercept  the channeled   beam and  to provide an online image of it  during data-taking. The detectors are placed in  a secondary vacuum vessel  that is separated from the primary vacuum by a 0.2 mm  thin aluminium layer. Such interface is 3.4 cm thick in the beam direction and could generate secondaries  when the vessel is moved into the beam.  
  
  A second Roman Pot   (at ``XRPH.51991'')  has been more recently installed;  it features four axes (two horizontal and two vertical) and it will be used in the next data taking sessions when detectors are installed inside its  four vessels.

   Few meters upstream the RP1 a LHC-phase-2 collimator is installed and used to cut 
  portion of the beam and derive measurements of the channeled beam.  
  
  In a position that is outside this   region  and  located  about 120 m downstream
  the crystals (at ``TAL.52196'', in Fig.\ref{fig:Layout2})    a 10 cm long  duralumin bar   is located  that is acting as a scraper. 
   This is a dispersive region where the Al bar  could  intercept off-momentum  particles that  are likely to  be generated in the interaction with the crystals.  Here the efficiency of the collimation system can be evaluated by detecting  the  scattered protons.      Scintillators and Cherenkov detectors are installed nearby to intercept the secondary particles  produced  by the protons inside  the Al bar.
  
    SPS was available to UA9 experiment during  five dedicated machine development   periods  
lasting  24--48 hours each during  the 2010.
   The machine was operated in coasting mode with  one single bunch that at the start of the
 fill contained about 10$^{11}$ protons.   One crystal at a time was  set in such  position such that it became  the primary collimator   and after that measurements could be performed whichÊÊincluded the angular scan of the crystal and scans with various absorbers.
  In stationary condition    protons from the  beam halo are diffusing into the edge of the 
crystal      with a rate such that around 100 protons  are reaching the crystal  at every turn  (23 
$\mu$s)  within the  bunch time length (3 ns).  This time structure of the halo protons is reproduced  in the signals
seen by the various detectors.

%Different regimes are in
%fact possible depending on the relative orientation of particle
%direction and crystal symmetry axes or planes: multiple
%scattering when a random orientation is chosen, axial channeling
%when particles are parallel to a line of atoms, volume
%reflection (VR) when the particles reverse their transverse
%momentum hitting the lattice planes, or the planar channeling
%- preferred in this experiment - when the particles are
%trapped between the lattice planes largely decreasing their
%interaction with the material. If the lattice planes are mechanically
%bent the channeled particles are deflected. Such
%configuration can be used as a primary collimator for a circulating
%beam instead of a more traditional passive scatterer
%[2].

\subsection{\label{subsec:cry} Crystals }

\begin{figure}[ht]
\centering
\includegraphics[width=0.5\linewidth]{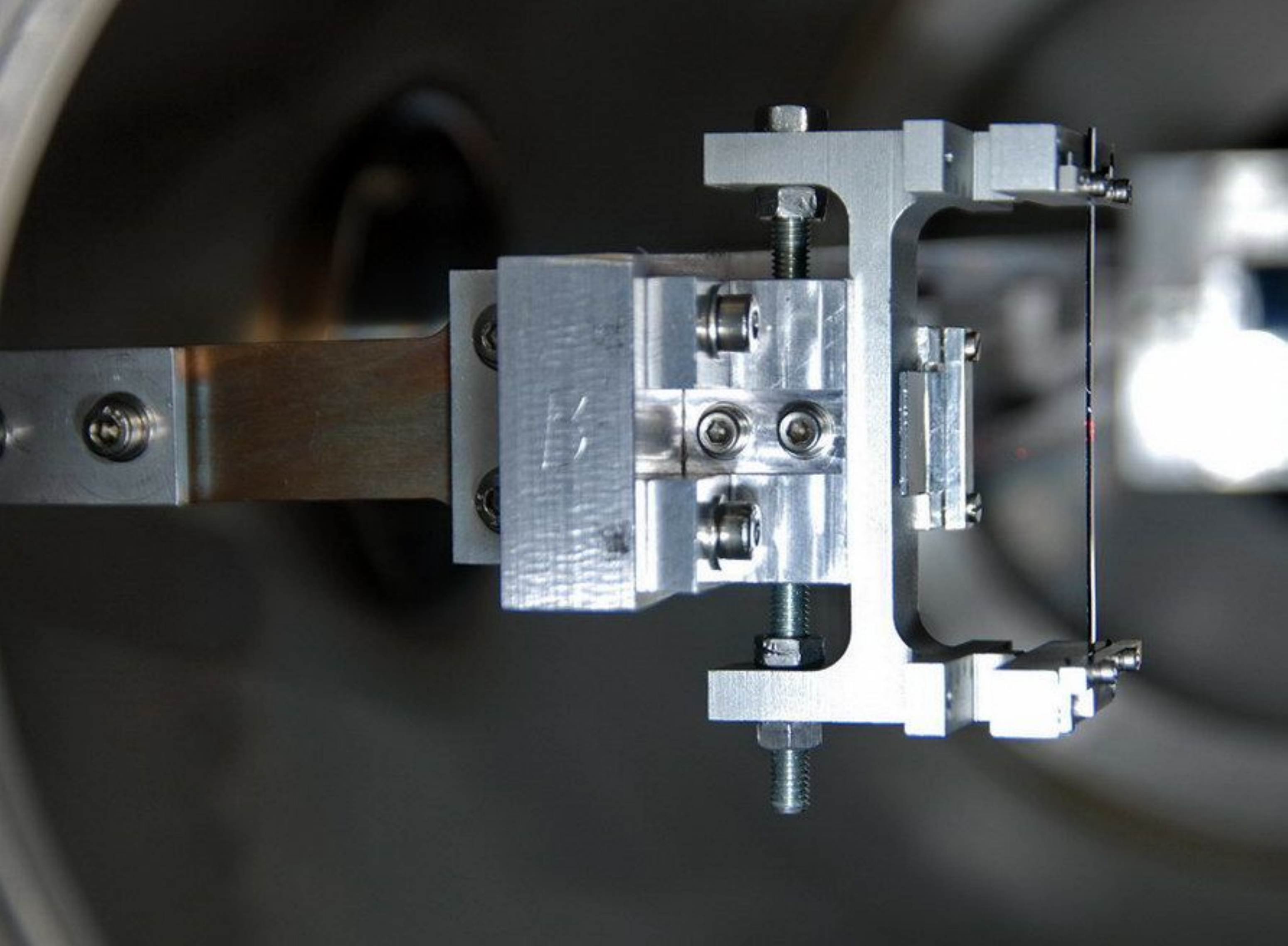}
\caption{Cr1 strip crystal in CERN  tank.}
\label{fig:stripFE}
\end{figure}

\begin{figure}[ht]
\centering
\includegraphics[width=0.5\linewidth]{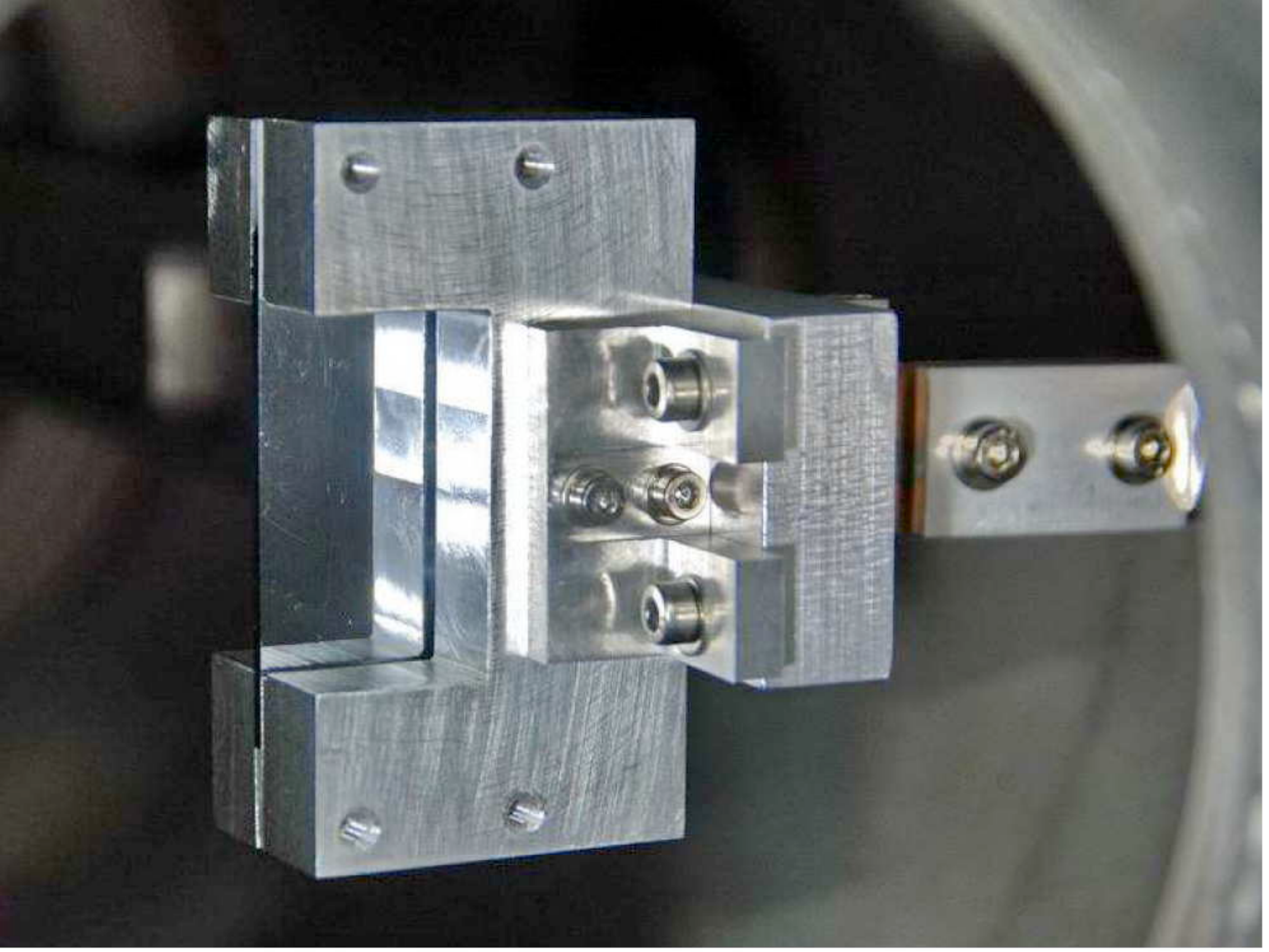}
\caption{Cr3  quasimosaic  crystal in IHEP  tank.}
\label{fig:quasiIHEP}
\end{figure}

The strip-like crystal Cr1  (Fig.\ref{fig:stripFE}) as well as  Cr4 has dimensions
$0.5\times70\times2$ mm$^3$ (width $\times$ height $\times$ thickness). They fabricated  at INFN Ferrara through micromachining techniques with the (110) planes parallel to the largest surfaces \cite{cr1cr4manufact}. A primary curvature is imparted to bend the strip crystal achieved by the  holder design, which result in a secondary curvature due to anticlastic deformation. This secondary deformation is used to steer incoming particles along the 2 mm size. Usage of anticlastic deformation enables to attain a highly uniform bending and to set the metallic components of the holder far from centre of the crystal and in turn far from the beam. 
%The ratio between the anticlastic over the primary curvature radii is modulated by anisotropy of Si and was recently calculated and measured to be 6.2 for the orientation under consideration [3]. The secondary deformation is used to steer incoming particles along the 2 mm size. 
The crystal Cr1 has been largely used for channeling and volume reflection studies \cite{cr1cr4H8} by the UA9 collaboration and top performed in efficiency (83\%) with a low-divergence beam. Bending angle of Cr1 crystal have been characterized by means of white light interferometry (Veeco NT1100) and resulted to be 150 $\mu$rad, while miscut angle have been measured by means of a Panalytical  high resolution X-ray diffractometer and resulted to be about 150 $\mu$rad.  Over a thickness of 2 mm a miscut angle of 100 $\mu$rad generates a region of reduced channeling efficiency as wide as 0.2 $\mu$m.
For Cr4, the holder has been equipped with a mechanical system to compensate for torsion in the strip crystal thanks to a feedback method during the standard stage of quality check of the crystal at fixed-target experimental area \cite{holdercr4}. Crystal miscut resulted to be 200$\pm$20 $\mu$rad. Both crystal torsion and bending angle (176 $\mu$rad) have been measured through deflection under planar channeling regime of a 400 GeV protons beam, available at SPS-H8 line. Crystal torsion, arising as a consequence of holder mechanical imperfections, has been reduced to 0.6 $\mu$rad/mm.

The quasi-mosaic  crystals Cr2 and Cr3  (shown in Fig.\ref{fig:quasiIHEP})  have been fabricated by Petersburg Nuclear Physics Institute (PNPI). The crystals are deeply polished  with submicron abrasive and
slow etching: this  yields  a perfect plane crystalline surface covered by several
fine grooves. They have a larger transverse  section than  strip crystals but
similar thickness in the direction parallel to the beam ($18\times15\times1.6$
mm$^3$  and  $30.5\times57.5\times2.10$ mm$^3$, respectively). 
 Their bending angles  are measured with optical systems  to be  150 and 165
 $\mu$rad and their miscut angles are 43 $\mu$rad and 92 $\mu$rad respectively. They
 are installed in  rigid holder frames that leave only a fraction of their
 transverse section to be exposed to the beam ( $2\times10$ mm$^2$ and $20\times40$ mm$^2$
 respectively).  Torsional effects induced by the holders have been measured with
 optical system to be  in the range 2--5 $\mu$rad/mm  depending on the vertical position on the crystals.

\subsection{\label{subsec:gon} Goniometers }
\begin{figure}[ht]
\centering
\includegraphics[width=0.5\linewidth]{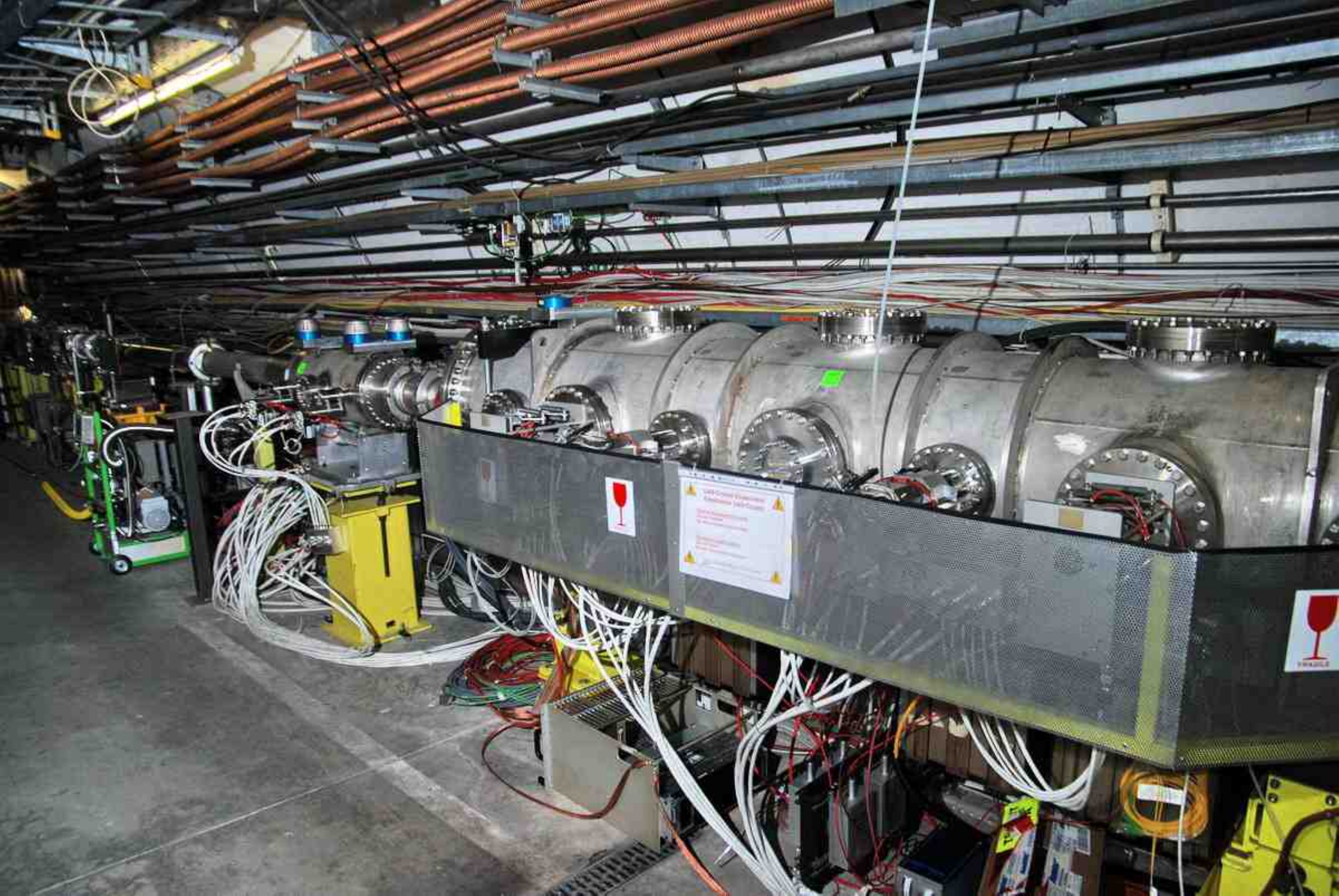}
\caption{IHEP and CERN tank.}
\label{fig:IHEPtank}
\end{figure}

\begin{figure}[ht]
\centering
\includegraphics[width=0.5\linewidth]{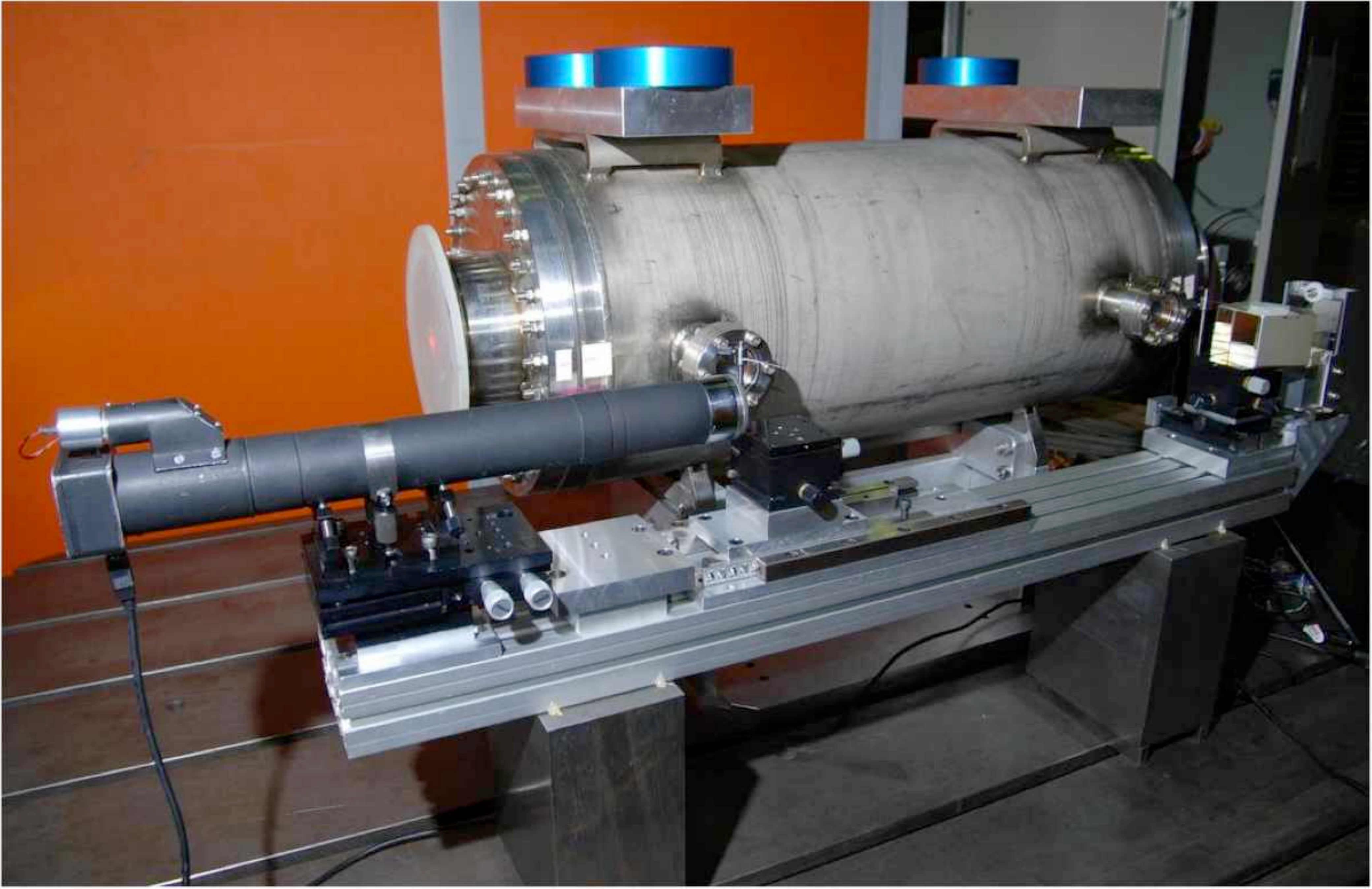}
\caption{IHEP   tank with autocollimator.}
\label{fig:IHEPautocoll}
\end{figure}

 In  the primary vacuum of CERN tank  at ``TEC 51795'' two high precision goniometers are installed. They allow a linear movement perpendicular to the beam in order to  insert the crystals into the beam   and   they generate a rotation in the horizontal plane. Crystals Cr1 and Cr2 are installed on these two goniometers.    In the IHEP tank at ``TECS.51793'' -  that is located  3 m upstream the CERN tank  - another mechanical system to which Cr3 and Cr4 
are connected is placed in the primary vacuum. In this case the crystals  are mounted on two mechanically connected supports that allow horizontal linear movement of each crystal: when one of the crystals is placed at a fixed distance from the beam, it can be rotated by applying a linear movement to the other support.  In Fig.\ref{fig:IHEPtank} the series of the two tanks is showed.
Both systems can rotate the crystal in angular range of tens of mrad. They  were designed  to have a resolution  (minimum step  achievable by the motor) of 1 $\mu$rad  and an accuracy (precision to which the motor goes to a given angular position) close to 10 $\mu$rad but such performances  were only partially reached during data-taking.  
In all goniometers the linear position of a pushing system  (measured with LVDT) is transformed to the
crystal angle via a Ògear boxÒ with some error. All goniometers are 
different but  can be characterized by the ratio of the real angular step measured with an optical laser 
autocollimator (Fig.\ref{fig:IHEPautocoll})  to the value of the  step in  the
motion controller. Such calibration were done in laboratory
and directly in situ through optical windows that are available in the CERN and IHEP
tank.  During angular scan in which the crystal is rotated with a fixed angular
velocity (1--10  $\mu$rad/sec)  the angular step calculated from LVDT readings was compared with the average angular step over one second.  Average deviations  from linearity were  measured and accounted for with an ad-hoc model.  From the RMS  deviation from linearity  we estimate  an accuracy of 10 $\mu$rad over an angular range of 15 mrad.

\subsection{\label{subsec:abs} Absorber }
\begin{figure}[ht]
\centering
\includegraphics[width=0.5\linewidth]{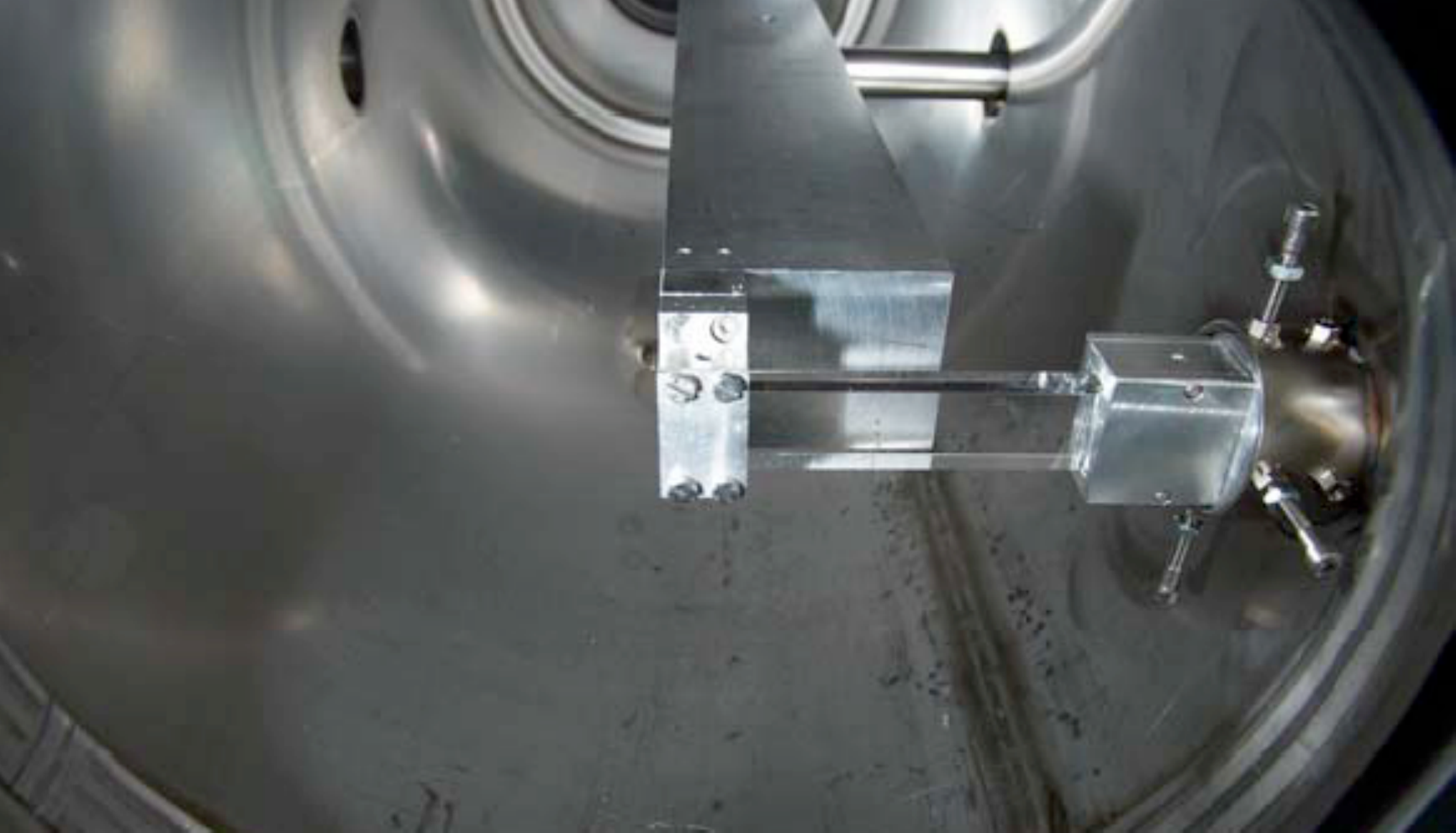}
\caption{Tungsten absorber at  TACW.51998  with quartz Cherenkov radiator.  }
\label{fig:absorber}
\end{figure}

 The absorber (at ``TACW.51998'') is a 60 cm long tungsten bar
 (Fig.\ref{fig:absorber}) with a  $7 \times 6$ cm$^2$ section located in a vacuum tank.  It can be moved  directly in and out of the beam and therefore   acts  as secondary collimator in all the measurements performed. Protons  generate showers in the material and they can leak out at a level of few per mille.  A  quartz crystal covering partially the bar entrance face generates Cherenkov light that is transported in the same crystals to a PMT. The light emitted is  proportional to the number of crossing protons and can be therefore used to measure the channeling efficiency in  this collimation system. The Cherenkov detector signal is acquired and digitized with an amplitude to frequency converter within a gate generated synchronously to the RF SPS signal. This is therefore  designed  to count protons originated from the single bunch  time structure of the beam.  Outside the vacuum tank scintillator detectors are installed which are sensitive to secondary particles produced in the hadronic showers and are therefore used to indirectly monitor the channeled beam.

\subsection{\label{subsec:LHCcoll} LHC-type  collimator}

The LHC-type collimator (``TCSM.51934'') is a full-scale prototype of the LHC  Phase II secondary collimator and has two horizontal, one-meter long copper ÒjawsÓ. The position and longitudinal tilt angle of each jaw can be controlled independently by means of four precise stepping motors (two per jaw) with minimum step size of 5  $\mu$m. The jaw positions are calibrated with respect to the nominal beam center and can be moved across the beam by 5 mm. The maximum collimator gap with fully retracted jaw is 60 mm. The two-sided collimator design allows to precisely define the beam envelope by closing the jaws around the circulating beam orbit and to identify its center. The collimator was also used to scan of the beam deflected by the crystal, in order to measure the channelling efficiency. The protons interacting in the jaws are producing secondaries that 
are  seen by LHC-type beam loss  monitors located downstream.

\section{\label{sec:BLM}  Beam loss monitors }

    Protons circulating in the accelerator hit obstacles and are therefore  either deflected, or lose  part of their energy or
    undergo  inelastic interactions  with nuclei. Detectors sensitive to showers of secondaries       produced in such interactions are  placed  along the beam line to measure the rate of such interaction.
    We generically indicate them  as beam loss monitor (BLM).      Different type of  detectors were used with different sensitivity to various range of interaction rates.

\subsection{\label{subsec:scint} Scintillators }
Polystirene  scintillator slabs with different sizes are mounted   outside the
beampipe on mechanical supports at critical locations of the layout.   They have an
approximately  $10\times10$ cm$^2$  square shape 
 with a thickness of 0.5--1.5 cm. They are readout with conventional photomultipliers (PMT)  connected to them via a light guide. Six of them  are  installed in pairs on the same support
 in a way coincidence of signals  from the two PMTs can be used to suppress low energy background.  Two more single scintillator counters are  used to monitor indirectly the beam losses. Those scintillators  are mostly sensitive to the charged component of the hadronic shower originating from 
 proton interaction in either crystals or collimator jaws.  During 2010 data-taking the PMT signals were  discriminated with a relatively high threshold  corresponding to several minimum ionizing particles (mips) energy release. The discriminated signals were fed into scaler's channels to count hits in a gate  of 20 ms fixed length.  The data acquisition system was realized with standard VME protocol in a Labview framework. 
  In the proximity of crystals scintillators  counting rates in the  crystal
  amorphous orientation were in the range of 1--10 KHz 
  depending of machine fill conditions. Given the time structure of the beam (one single bunch) several protons are hitting the crystals or other obstacles within few ns. Since no pulse height information is retained scintillators are not able to separate the contribution  of the single protons and therefore their counting  rate tend to  saturate at the machine revolution frequency (43 KHz).  
  They were especially  useful  during the online data analysis  to find the crystals channeling condition  and in offline analysis to determine the beam loss patterns in channeling versus amorphous condition. 
 
 \subsection{\label{subsec:PEP} PEP-II-type detectors }

 The PEP-II  beam-loss monitors  \cite{PEPdet}  are   detecting Cherenkov light using a 16-mm photomultiplier with about 2 ns intrinsic pulse width.
 %  (a few 10s of  % longer in the UA9 installation due to dispersion in the long signal cables). 
  The radiator is an 8 mm diameter, 10 mm long fused-silica cylinder placed against the fused-silica PMT window with optical grease on the interface. The opposite end and the cylindrical surface are aluminized for high reflectivity. The assembly is enclosed in 10 mm of lead, originally provided to avoid synchrotron-radiation background in the PEP-II application. The outer shell is a magnetic steel cylinder to provide some shielding against stray magnetic fields. In UA9 the detectors are run at 800 to 900V on the PMT and into a discriminator set to 15 mV. The dark rate is a few counts/min under these conditions.
 Three detectors of this type  were installed along the beam line, a pair of them in the proximity of a SPS collimator  (not used during 2010 test)  and another close to the tungsten absorber region to monitor the losses 
 downstream the RP1 and the LHC collimator.

\subsection{\label{subsec:GEM} GEM chambers }
A  triple GEM   detector is a micro-pattern gas detector
which consists of a primary ionization gap and three consecutive GEM foils \cite{sauli}.
A printed circuit board with readout pads detects the current induced by the drifting electron cloud
originating from the last GEM stage. Thus the gas amplification and the signal readout are completely
separated. 
 The detectors used in SPS layout  are built starting form the standard GEM foils produced by CERN with 10$\times$10 cm$^2$ of active area. 
The anode PCB 12$\times$12 cm$^2$ has been designed to  house 128 $6\times12$-mm$^2$ pads inside the active area,
while keeping eight connectors for the front end electronics (FEE)  in the opposite side placed always in the same position. The three foils  are successively glued together forming four gaps
(3, 1, 2, 1 mm), following the same structure used for the LHCb chambers \cite{alfonsi}.
 The FEE boards used for this development are based on Carioca-GEM Chip \cite{bonivento} 
 (used for LHCb). The boards, designed and realized at  LNF,  house 16 ASD channels (2 chips) that produce LVDS time over threshold signals.
A mother board finally is plugged just on top of the eight boards for low voltage power supply, threshold distribution, and high voltage filters.
The LV supply of 2.5 V and the threshold settings are driven through a NIM module that can be set manually or remotely, through
a VME programmable DAC (0$\div$2 V),   while the HV is supplied  through a  module\cite{corradi} developed at LNF  that  starting from a 12 V voltage
  produces  the three GEM voltages (0$\div$500 V) determining the chamber gain, and the four voltages (0$\div$1200 V) producing the electric fields for electron cascade drift.
   The chambers are flushed with a gas mixture $Ar:CO_2:CF_4$ ($45:15:40$).
  
\begin{figure}[ht]
\centering
\includegraphics[width=0.5\linewidth]{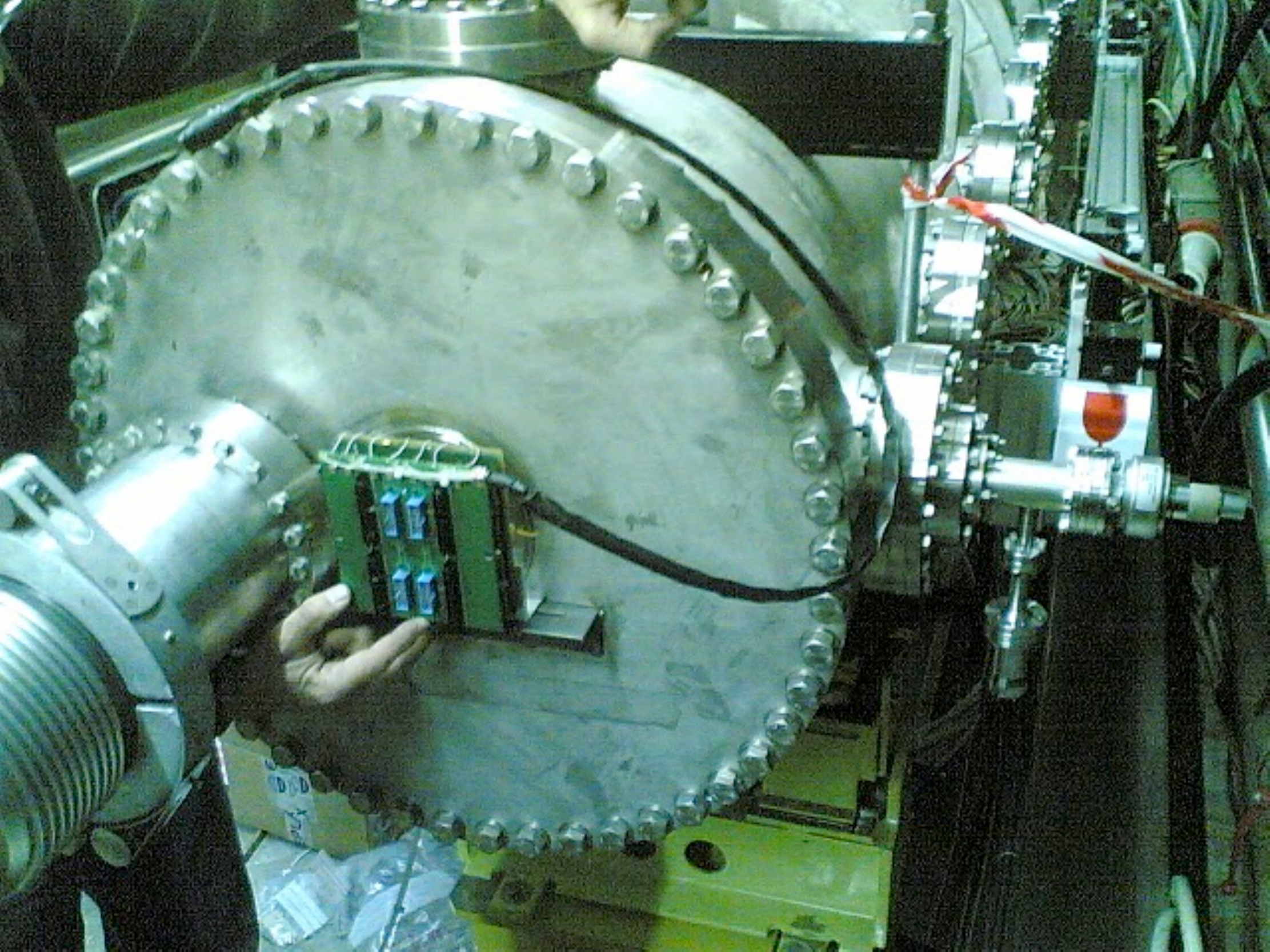}
\caption{GEM detector during installation on CERN tank (downstream side).}
\label{fig:GEM}
\end{figure}

   Two identical  triple GEM  detectors are installed on the external walls  of the CERN tank (Fig.\ref{fig:GEM}),  close to the beam pipe with the anode pads perpendicular to the beam direction.   They are therefore sensitive to the secondaries produced in the IHEP tank and in the CERN tank. Those detectors are virtually  able to count each single charged particle crossing the sensitive volume.  Given the relatively small size of each pad,  thresholds can be    set at the one mip level.
   % In Fig. \ref{fig:GEMratio} the ratio of the counts in amorphous over the counts in channeling position are given for Cr3 showing... 
    Single pad counting rates in amorphous condition were about  1 KHz.  Given the lower threshold the  GEM  detectors were   complementary to the scintillators, being more efficient to count particles with relatively lower rates.
       Anode current readings  proportional to the hit rate are  used in online monitoring during data-taking and in offline analysis to compare with rates measured by other BLM.

%  \begin{figure}[ht]
%\centering
%\includegraphics[width=0.99\linewidth]{GEMratio.pdf}
%\caption{.}
%\label{fig:GEMratio}
%\end{figure}
%

\subsection{\label{subsec:Medipix} Medipix}
\begin{figure}[ht]
\centering
\includegraphics[width=0.5\linewidth]{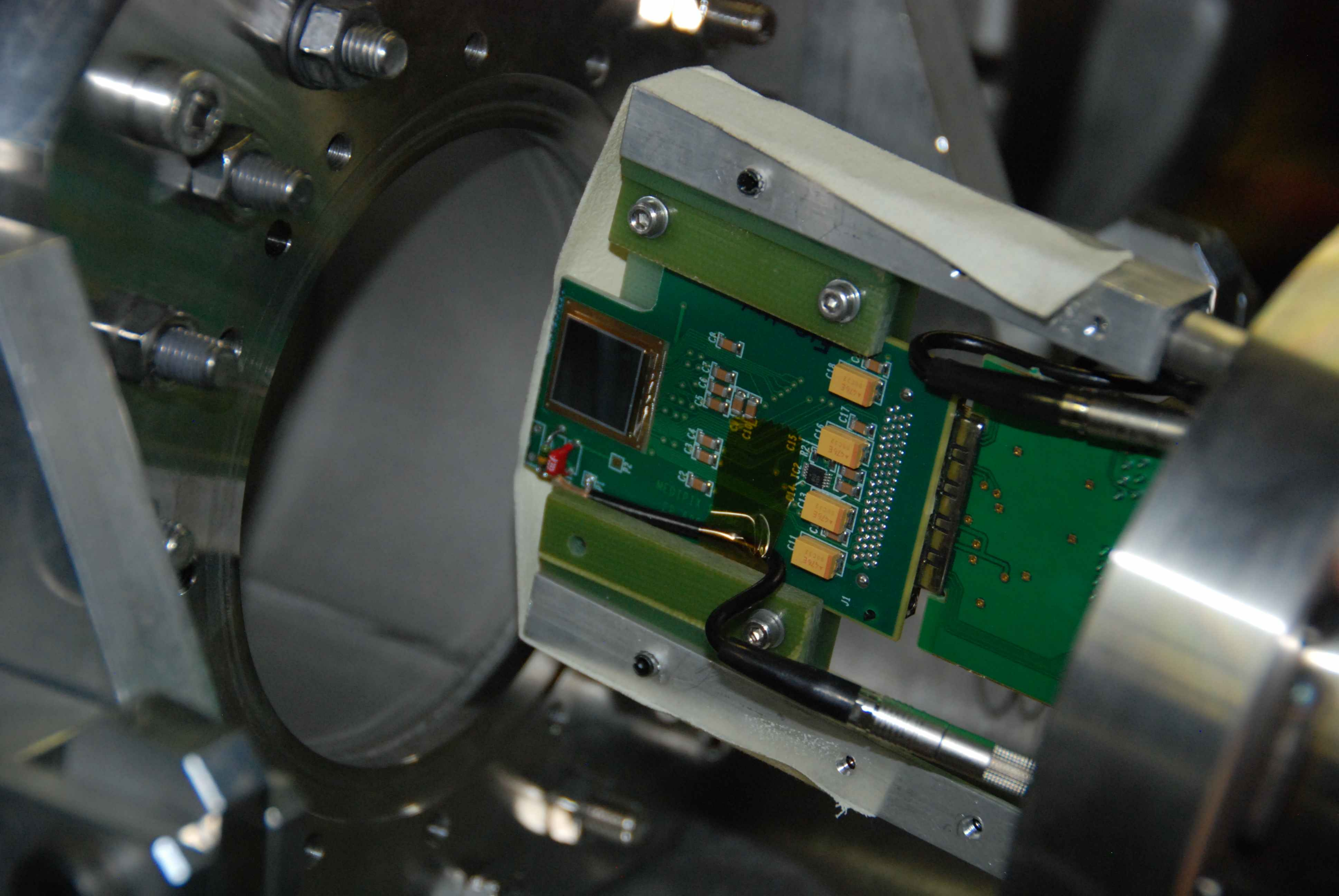}
\caption{MEDIPIX sensor and readout card to be inserted in RP1.}
\label{fig:Medipix}
\end{figure}

\begin{figure}[ht]
\centering
\includegraphics[width=0.5\linewidth]{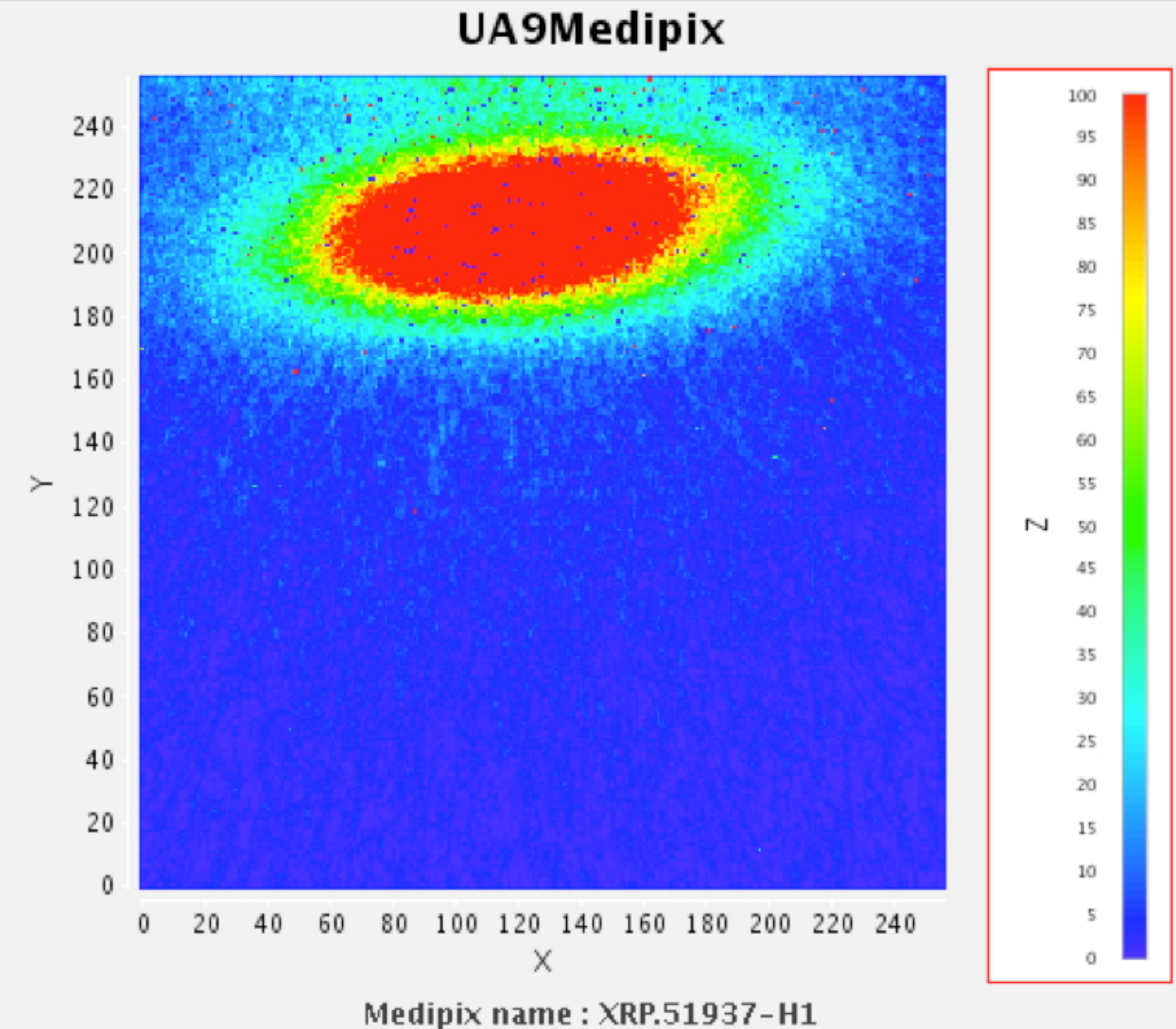}
\caption{Online Medipix image of the channeled beam. The Y axis corresponds the relative position of the edge of the sensor along the radial direction in the machine horizontal plane. The  X axis corresponds to the relative vertical position of the sensor. For each pixel the number of counts   (proportional to the number of crossing protons ) in a gate  1 s long is reported.  }
\label{fig:MedipixCR1}
\end{figure}

The Medipix2 MXR ASIC\cite{medipixpaper} is a high spatial, high contrast resolving CMOS pixel read-out chip working in single photon counting mode. It  has originally been designed to be combined with different semiconductor sensors  segmented into pixels to detect   X-rays and used for  X-ray and gamma-ray imaging applications. In UA9 it was used to detect the ionization charge released  by the proton crossing the silicon pixels.  The preamplified signal from each pixel  is discriminated to generate one count. The threshold can be adjusted pixelwise with 4 bits for uniform performance of the whole pixel matrix. Counts are accumulated during a predefined exposure time that can be chosen between few tens $\mu$s to several seconds and data are stored  in  13-bit counter per pixel. Read-out is performed at the end of the exposure to avoid dead time.
 In each of the  two arms of RP1 a  Medipix detector  was  installed and used to
 intercept directly the channeled beam. They have  a $1.4\times1.4$ cm$^2$ active
 area segmented in  $55\times55$-$\mu$m$^2$ $256\times256$ pixels. To avoid dead space the board housing the sensor was cut and the edge of the detector put at 1.53 mm from the internal edge of  the  thin Al window closing the RP1 vacuum vessel (Fig.\ref{fig:Medipix}).
Energy calibration of such detectors  was done using H8 extracted proton beam and
sparse data frames (10-$\mu$s long) with crystal in channeling condition. The mean number of counts per hit is  about 1.5 with a large uncertainty (20\%).   In Fig.\ref{fig:MedipixCR1}  a  Medipix image with Cr1 in channeling condition for the inner Medipix installed in Roman Pot 1 is shown.  The frame exposure time was set at 1 sec. The channeled beam is clearly visible and this guided online data-taking. Offline analysis used these frames to compare with beam lifetime to deduce  the  extraction efficiency\cite{medipixlaface}.
  Besides the two detector in RP1, a  third Medipix2 detector has been installed outside the beam pipe in order  to monitor beam losses in a region downstream the crystals.

\subsection{\label{subsec:LHCBLM} LHC-type BLMs }

LHC-type beam loss monitors\cite{LHCBLM}   are  ionization chambers with parallel
aluminum electrodes separated by 0.5 cm. The detectors are about 50 cm long with a
diameter of 9 cm and a sensitive volume of 1.5 liter. The chambers are filled with $N_2$ at 100 mbar overpressure and operated at 1.5 kV. Their  counts are  integrated over a  1.2 s period.

Eight LHC BLM were installed in various positions of UA9 experimental region. They were especially useful to detect condition of losses characterized by   high rates (100 KHz- 1MHz). When  UA9 scintillators tend to saturate,  LHC BLM  are giving a linear response. On the contrary for relatively low rate  (few KHz)  they do not provide  precise measurements.

\subsection{\label{subsec:SPSdevice} SPS devices  }

   Several SPS devices would be available to the UA9 experiment, but they were generally  not sensitive to the relatively  low current used in the experiment.
   Some of them were anyway  used to measure  the beam current and  deduce the beam lifetime.
    BCT   were pick-up device detecting an  induced current due to the passage of
    the beam particles. The current is integrated over  a 10   ms gate and every 16--18 s the information is readout. 
    The SPS current is derived summing the charge information over 1 sec. From time
    derivative  of this measurement  the total loss of protons can be derived with
    some large uncertainty (20--30\%).
  Morevoer,     wire scanners were  used to measure the   emittance of the beam at each fill. 

\section{\label{sec:software}  Control software and data acquisition }

\begin{figure}[ht]
\centering
\includegraphics[width=0.8\linewidth]{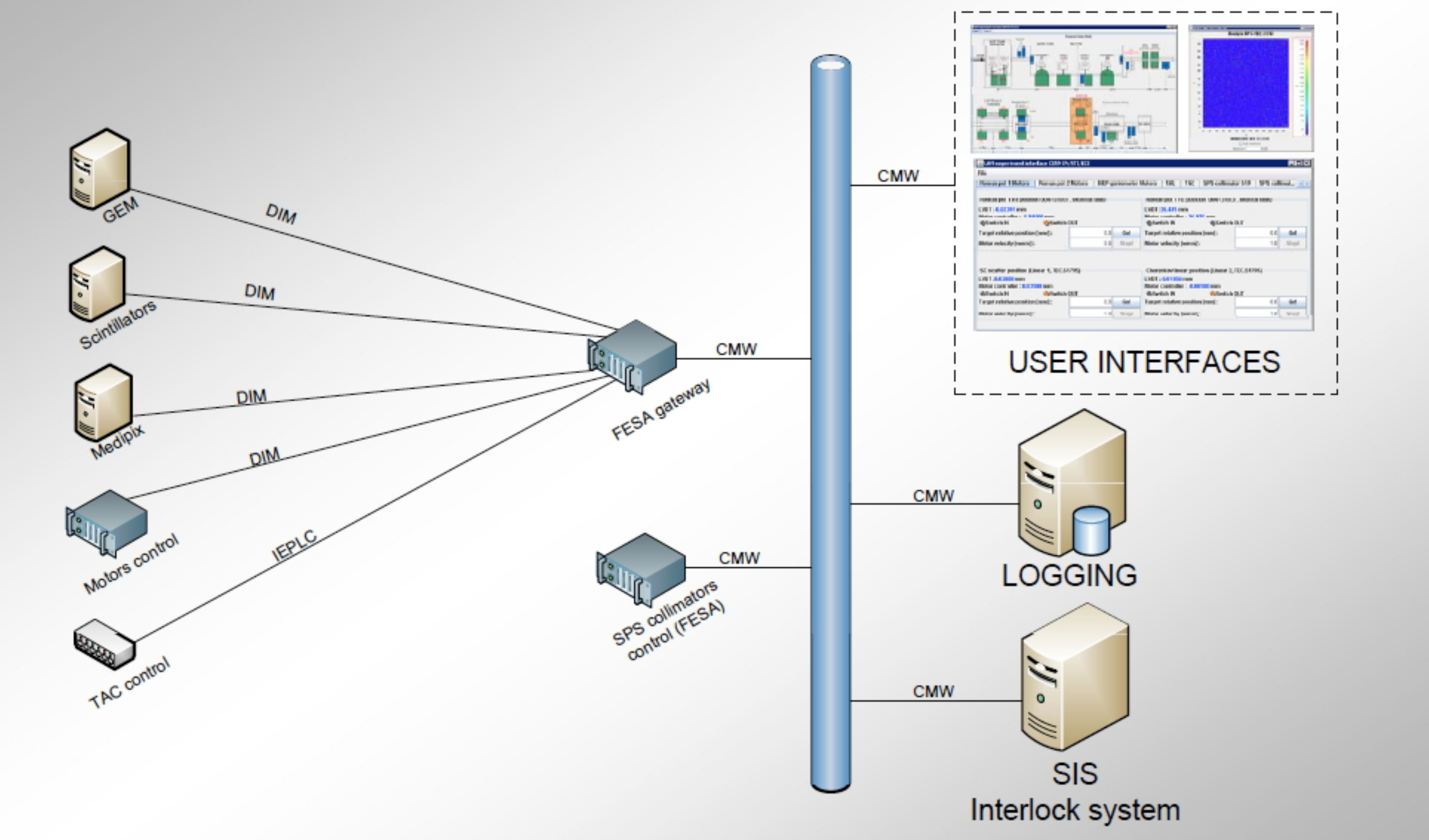}
\caption{Control and data acquisition scheme.}
\label{fig:controlsoftware}
\end{figure}

 All UA9 devices are controlled via common interface that is able to change positions of the movable devices, to monitor them and to record 
 the rate measurements of the various detectors (Fig.\ref{fig:controlsoftware}).  A  powerful middleware is therefore necessary to bring in a single environment data coming from both the
machine control system and the UA9 experimental devices, that have conceptually different low level control electronics.
 The motorization low-level control is based on Labview real-time PXI chassis that talk to the motor drivers or the LVDT transducers through FPGA cards.  All the data from the various devices are in fact coming via different systems and need to be unified and synchronized and then made public over the TCP/IP network (Fig.\ref{fig:controlsoftware}). A Distributed Information Management (DIM) system used by LHC experiment and capable to connect to different platform is used to collect and send data to various devices through a Front End System Acquisition Architecture  (FESA) gateway.  The user can connect to the control system via Java application; positions and detector rates are available online through a graphical interface.  Every minute the system saves the whole  information into text files and guarantees  the synchronization. 
 
  \section{\label{sec:concl} Summary}
    In this paper we have described the UA9  experimental layout that has been deployed to study the crystal based collimation in the  SPS coasting beam at CERN during 2010.
     Several high precision movable devices were operated in a vacuum environment to orient crystals in the channeling configuration and to check the deflection of the halo protons far from the primary beam.
      Beam loss monitors of various types were  used to detect secondary particles produced  in inelastic interaction of protons with the accelerator apertures. All this system has been operated in several tests
      in which the channeling condition was established in a very reproducible way.

\begin{acknowledgments}

Work supported by the EuCARD programme GA 227579, within the ``Collimators and Materials for high power beams'' work package (Colmat-WP8).
 G. Cavoto and R. Santacesaria acknowledge the support from MIUR (grant FIRB RBFR085M0L 001/I11J10000090001). Some part of this work supported by US DOE under the LARP framework.
\end{acknowledgments}

%\appendix

\end{document}